\documentclass[preprint]{revtex4-2}

\usepackage{graphicx}% Include figure files
\def \co322{Co$_3$Sn$_2$S$_2$}

\begin{document}

\title[]{Exchange biased Anomalous Hall Effect driven by frustration in a magnetic Kagome lattice}

\author{E. Lachman*$^{1,2}$, N. Maksimovic$^{1,2}$, R. Kealhofer$^{1,2}$, S. Haley$^{1,2}$, R. McDonald$^{3}$ and James G. Analytis$^{1,2}$}
 \altaffiliation{corresponding author: ellal@berkeley.edu}%Lines break automatically or can be 
 \affiliation{\\
$^1$Department of Physics, University of California, Berkeley, California 94720, USA\\
$^2$Materials Science Division, Lawrence Berkeley National Laboratory, Berkeley, California 94720, USA\\
$^3$Los Alamos National Laboratory, Los Alamos, New Mexico 87545, USA
}

\begin{abstract}
\co322\, is a ferromagnetic Weyl semimetal that has been the subject of intense scientific interest due to its large anomalous Hall effect. 
We show that the coupling of this material's topological properties to its magnetic texture leads to a strongly exchange biased anomalous Hall effect. We argue that this is likely caused by the coexistence of ferromagnetism and spin glass phases, the latter being driven by the geometric frustration intrinsic to the Kagome network of magnetic ions.

\end{abstract}

\maketitle

\section{\label{sec:intro}Introduction}

Magnetic Weyl semimetals (WSM) are predicted to host the Quantum anomalous Hall effect at higher temperatures than magnetically doped topological insulators \cite{Muechler2017}, and are therefore of substantial interest for spintronics technologies. One such material is \co322, which has been the subject of intense research interest because the interplay of topology and magnetic order leads to a giant anomalous Hall effect (AHE) in the presence of a weak ordered moment \cite{Liu_NatPhys_2018}. \co322 is a shandite material, where the Co atoms form layers of 2D Kagome lattice. S and Sn atoms are interleaved in the layers, with another Sn species in between the Co-S layers. \co322\, is a half-metallic material, with only one component of spin contributing to the conductivity in the ferromagnetic state below $T_c = 175$~K \cite{Schnelle2013}. More recent calculations and measurements including ARPES and STM show that \co322\, is a WSM, with Weyl points located $\sim60$~meV above the Fermi energy \cite{Morali2019, Yin_NatPhys_2019}. However, the magnetic order is not straight forward, and several studies have recently suggested this material hosts a more complex magnetic texture.
Understanding how this magnetic complexity affects the topological properties of \co322, and specifically the AHE, is the focus of this study. 

Exchange Bias (EB) is a property that is very important in magnetic memory technologies, ensuring stability and protecting against volatility. Typically, it is the result of exchange interaction at the interface of a ferromagnet (FM) and another magnetic phase, typically an anti-ferromagnet (AFM) \cite{Meiklejohn1957}. As a result of this interaction, the EB effect manifests as a shift of the magnetic hysteresis loop. This shift is related to the strength of the pinning exchange interaction between the FM and the AFM \cite{Nayak_nmat_2015}. Defining $H_{C\pm}$ as the field at which the magnetization changes sign along an $M(H)$ curve, the exchange bias can be parametrized as $H_{EB} = -(H_{C-} + H_{C+})/2$. Here we show that the coexistence of two magnetic phases in \co322\,leads to an exchange bias that strongly influences the AHE. In addition to being of fundamental interest, this opens the possibility of applying this mechanism as a basis for topological spin valve technologies.

\section{\label{sec:eb} Experiment and Results}

Single crystals of \co322 were grown using the flux method (see methods for details) and were either used pristine or silver epoxy contacts were attached for transport measurements. A plot of resistance as a function of temperature presented in Fig.~\ref{fig:cooldownAndEB}(a) indeed shows a ferromagnetic transition at a temperature of $175$~K. 

\subsection{Low temperature anomalous Hall effect and Exchange bias}

The magnetic easy axis in \co322\, is perpendicular to the Co Kagome planes. The system exhibits a magnetic hysteresis loop when sweeping a perpendicular magnetic field at temperatures below the magnetic transition.

\begin{figure}[ht]
    \centering
    \includegraphics[width = 0.48\textwidth]{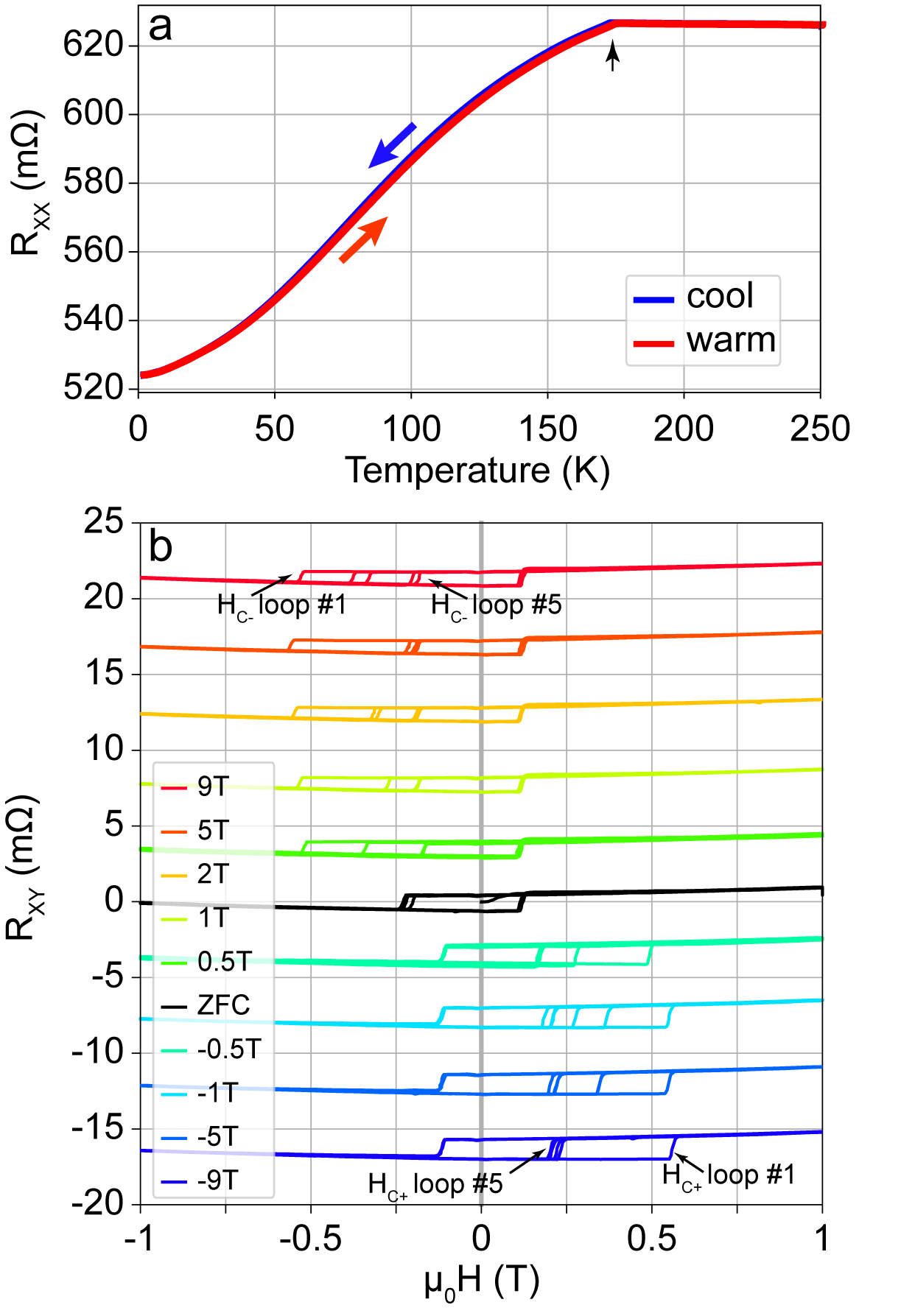}
    \caption{Magnetic transition and low temperature behaviour of \co322\, single crystal sample. (a) Longitudinal ($R_{xx}$) resistance as a function of temperature shows a magnetic phase transition at $175$~K as evident by the change in slope (marked with black arrow).
    (b) Exchange biased AHE. Hall resistance as a function of magnetic field at a temperature of $2$~K. The field is swept from between $+1$~T and $-1$~T and the sweep is repeated 5 times, resulting in a different $H_{C-}$ ($H_{C+}$) for cooling down in positive (negative) magnetic fields. The coercive fields on the opposite side are the same and therefore fall on top of each other in the plot.}
    \label{fig:cooldownAndEB}
\end{figure}

At a temperature of $2$~K, the magnetic transition appears as a sharp step in $R_{xy}$ at $\mu_0 H_c = 115 \pm 10$~mT (Fig.~\ref{fig:cooldownAndEB}(b)).
When accounting for mixing of the longitudinal resistance (see methods), the resulting hysteresis loop is characteristic of the AHE with a crucial difference: The loop is not centered around zero applied field with the offset depending upon the thermal and magnetic history. The transition at positive applied fields is at $\mu_0 H_c = 115$~mT, but on the negative part it is between $\mu_0 H = -200$~mT and $\mu_0 H= -230$~mT and changes between sequential repeated field sweeps.
When cooling the sample in the presence of magnetic fields, a negative field shifts the positive coercive field as far as $550$~mT initially, and to $200$~mT after relaxation with repeated field sweeps. Positive fields shift the negative coercive field to similar fields with opposite sign. This asymmetry of the hysteresis loop is the signature of EB \cite{Meiklejohn1957,RLStamps2000}, and serves as evidence that the interpretation of the magnetic phase of \co322\, as a simple FM phase in which the $Co$ spins point out of plane is lacking.

\begin{figure}[ht]
    \centering
    \includegraphics[width = 0.48\textwidth]{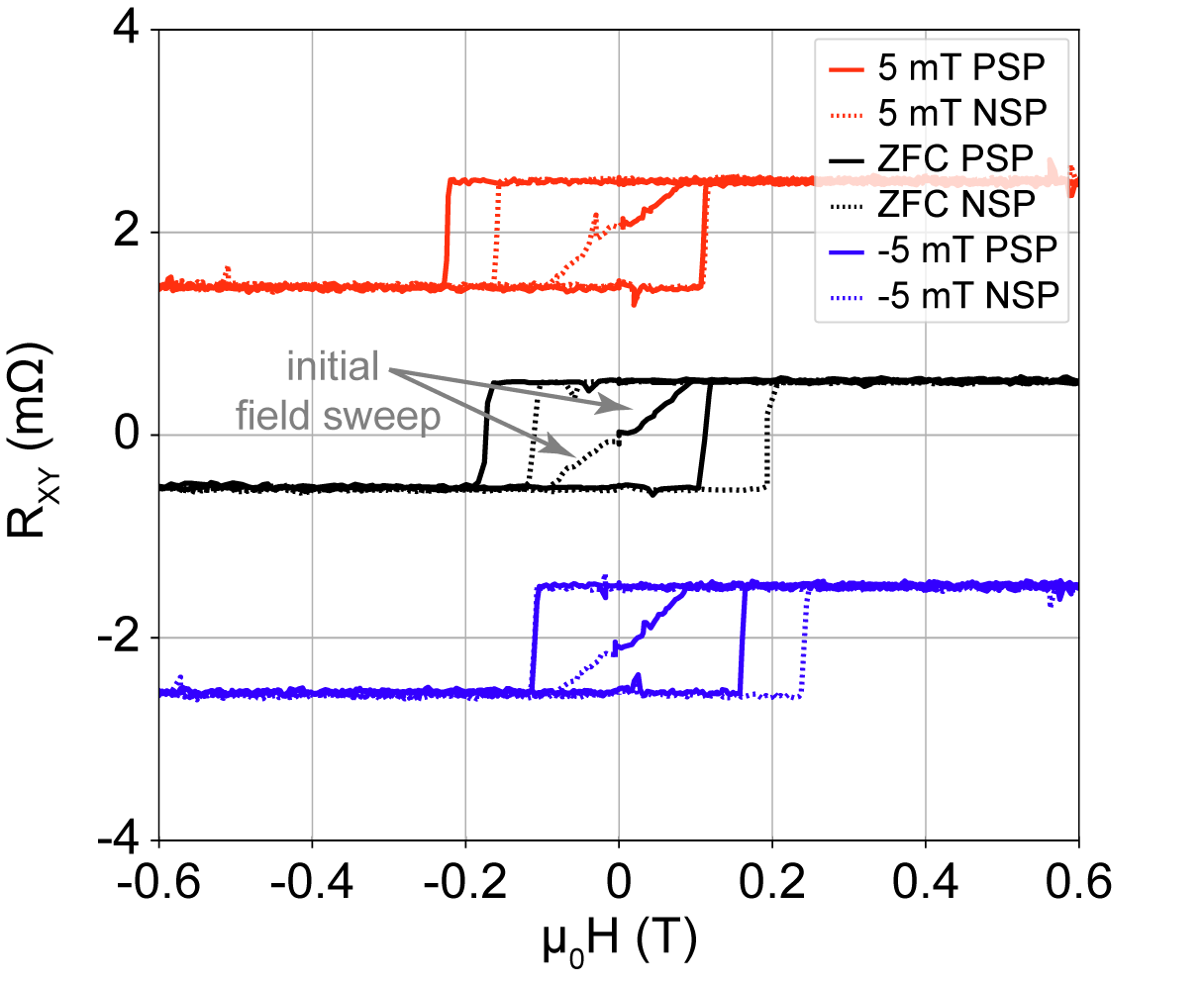}
    \caption{Spontaneous exchange bias. $R_{xy}$ as a function of applied magnetic field for different low values of cooling fields at $2$~K. For each field, both a positive sweep protocol (PSP) and a negative sweep protocol (NSP) were performed, where the sample is warmed up to $250$~K - well above $T_c = 175$~K - between the two to negate the SEB magnetization effect. The initial field sweeps for the zero field cooled curve are marked with arrows, but are clear inside the hysteresis loop area for all fields. These indicate that the initial magnetic state is unsaturated, and that the SEB is indeed induced at the low temperature, by the specifics of the field sweep protocol.}
    \label{fig:seb}
\end{figure}

Remarkably, even when cooling the sample with no magnetic field applied (zero-field cool - ZFC) a small EB appears (see Fig.~\ref{fig:seb}, this can be seen in Fig.~\ref{fig:cooldownAndEB}(b)). Though first attributed to a possible remnant field in the superconducting magnet, further investigation shows this EB to be a spontaneous one \cite{Wang_PRL_2011}. The spontaneous EB (SEB) can be isothermally induced by the initial field sweep direction at low temperatures, and demonstrates the importance of the specifics of the field sweep protocol. Figure~\ref{fig:seb} shows magnetic field sweeps at $2$~K, with two different field sweep protocols. One is ``positive'' where the field is swept $0 \rightarrow +1T \rightarrow -1T \rightarrow +1T$, and the other is a ``negative'' sweep protocol, where the field is swept $0 \rightarrow -1T \rightarrow +1T \rightarrow -1T$. These two sweeps were each performed after zero field cooling and after in-field cooling. Let us first analyze the meaning of the zero field cooled sweeps. The positive sweep protocol (PSP) results in $H_{EB} = 31.5$~mT, and the negative sweep protocol (NSP) yields $H_{EB} = -40$~mT. Both positive and negative protocols produced the same saturation value of $\Delta R_{xy} = 1 m\Omega$, and show a similar magnitude of $H_{EB}$. In addition, the initial value of $R_{xy}$ for both field sweeps is close to zero, indicating that the initial state of the material is unmagnetized or of very low magnetization. This rules out the naive explanation of remnant field in our magnet.
When the sample is cooled in a small field, the SEB effect and the normal EB effect combine to form the total EB. This is evident, for example, in the $5$~mT cooling PSP, in which $|H_{C-}|$ is larger than that of the NSP. The same applies for the $-5$~mT cooling, but now it is the NSP in which $H_{C+}$ is larger than that of the PSP.
The presence of EB and SEB at low temperatures would suggest that an additional phase or interaction of an AFM or spin glass (SG) nature as well as the FM is also preset in \co322 \cite{Meiklejohn1957, Ali_NatMat_2007}.

\subsection{\label{sec:125K} Anomalous Hall effect and magnetic properties at intermediate temperatures}

\begin{figure}[ht]
    \centering
    \includegraphics[width = 0.5\textwidth]{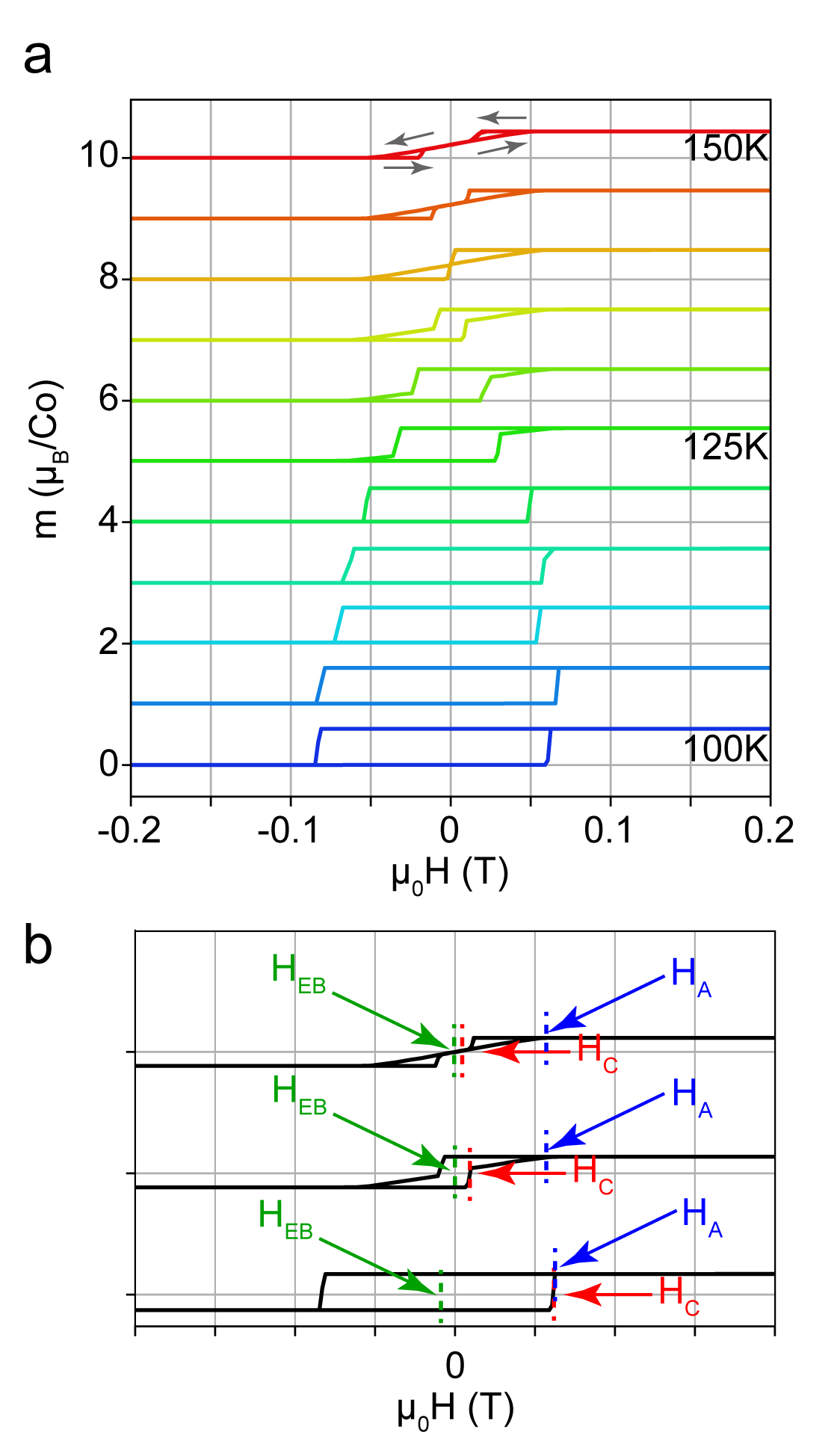}
    \caption{Magnetic hysteresis at intermediate temperatures with field parallel to the c-axis.
    (a) Magnetization as a function of applied magnetic field for a sample cooled in an applied field of $0.5$~T showing the magnetic hysteresis loops for temperatures near $125$~K. The square shape of the loop at $100$~K gradually changes to a bipartite transition at $125$~K.
    (b) Schematic definition of the anisotropy field ($H_A$), the coercive field ($H_C$) and the exchange bias field ($H_{EB}$) on the three types of loop shapes. For the low temperature square loops, $H_A$ and $H_C$ coincide.
    }
    \label{fig:biasAtdiffFieldsTemps}
\end{figure}

In order to look for clues as to the nature of the EB, magnetization measurements as a function of magnetic field were performed at different temperatures for a sample cooled in a field $\mu_0 H_{cool} = 0.5$~T (Fig.~\ref{fig:biasAtdiffFieldsTemps}(a)).
These hysteresis loops reveal a qualitative difference between low temperatures and high temperatures still below the Curie temperature of $175$~K. A clear change in the shape of the hysteresis loop appears at $125$~K, and the coercive field (where $m$ crosses zero) is reduced significantly above this temperature. The unconventional shape of the hysteresis loops above $125$~K suggests an unusual energy landscape for the phase underlying the EB, and some interplay between it and the ferromagnetism.  

\begin{figure}[ht]
    \centering
    \includegraphics[width = 0.7\textwidth]{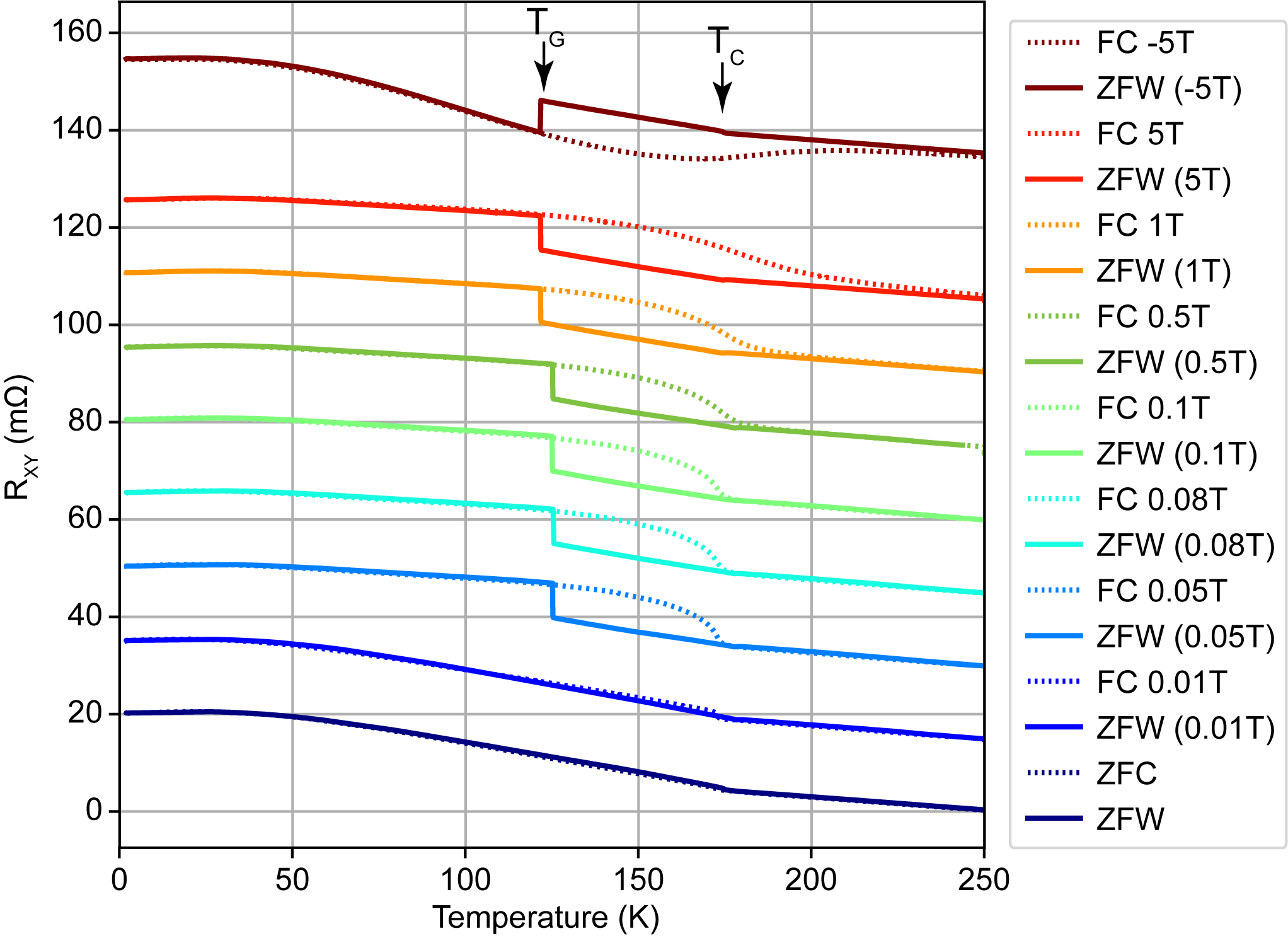}
    \caption{A new phase transition at $125$~K.
    (a) $R_{xy}$ as function of temperature for zero field warming up (ZFW) after in-field cooling (FC) of the sample. For each FC (dotted lines) the ZFW (solid) is of the same color. On the ZFW curves, the $125$~K transition is evident as a sudden change in $R_{xy}$ to a different value. For cooling in positive (negative) field, the change in $R_{xy}$ at $125$~K is to a lower (higher) value.
    }
    \label{fig:warmups}
\end{figure}

When looking at the sample's resistance as a function of temperature in Fig.~\ref{fig:cooldownAndEB}(a), there is no clear signature of an additional magnetic phase transition that may explain the appearance of EB. However, the existence of a transition at $125$~K becomes readily apparent when studying the effect of an applied field on $R_{xy}$, as shown in Fig.~\ref{fig:warmups}. A field was applied while cooling (where FC data is collected), swept to zero at $2$~K and then the sample was warmed up to $250$~K (where the ZFW data is collected). For cooling fields $\vert \mu_0H_{cool} \vert > 0.05$~T, a clear transition can be observed at $125$~K, marked by a sudden change in $R_{xy}$ in the ZFW curves. We denote this temperature as $T_G$. This demonstrates there is an interaction between the FM and another phase, which pins the zero-field $R_{xy}$ below $T_G = 125$~K.
FC and ZFW curves merge at the FM transition at $175$K. This suggests that the phase below $T_G$ retains the memory of the cooling conditions, and forces a change in the FM ordering at zero field.

\subsection{Magnetic and thermodynamic properties of the $125$~K phase}
\begin{figure*}[ht]
    \centering
    \includegraphics[width = 0.95\textwidth]{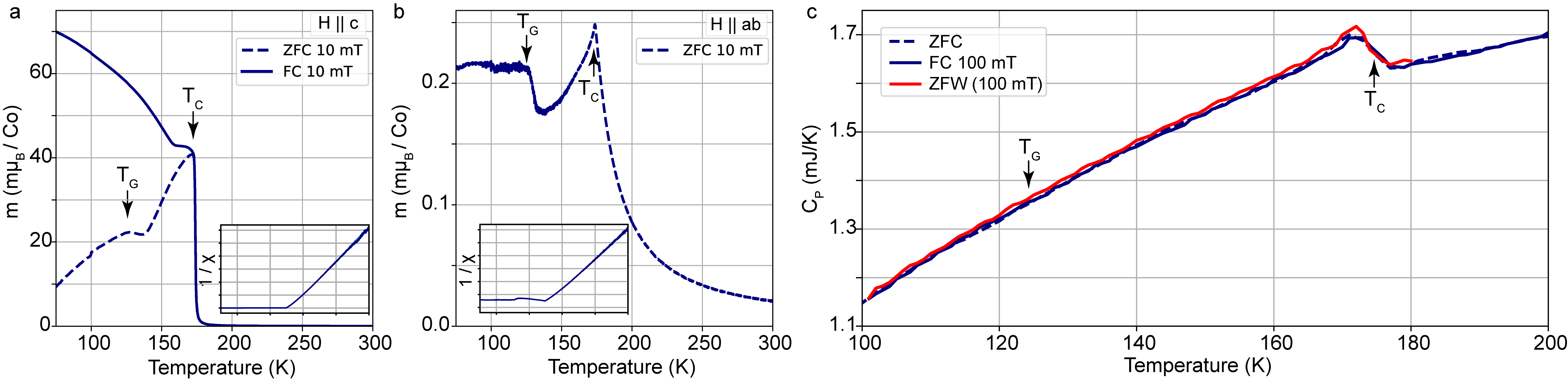}
    \caption{(a+b) In-plane and out-of-plane magnetization measurements on a single crystal. 
    (a) Out-of-plane (H $||$ c) magnetization as a function of temperature while warming up the sample after ZFC (dashed) and field cooled (solid) and measuring under a field of $10$~mT. The curves are characteristic of a FM with a Curie temperature of $175$~K, with a small feature at $125$~K (marked with a small arrow on the ZFC curve).
    (b) In-plane (H $||$ ab) magnetization as a function of temperature while warming up the sample after ZFC under a field of $10$~mT. The curve is characteristic of an AFM with a N\'eel temperature of $175$~K, with a feature at $125$~K.
    (c) Single crystal heat capacity as a function of temperature for $100-200$~K. These measurements reveal no feature at $125$~K, despite repeating the ZFW after FC protocol as in Fig.~\ref{fig:warmups}.}
    \label{fig:magnetization}
\end{figure*}

A recent work using muon spin rotation \cite{Guguchia_arXiv_2019} concluded that the magnetism in \co322\, originates solely from the Co atoms, and that there is a phase transition from a higher temperature in-plane AFM to a low temperature out-of-plane FM at $\sim90$~K. This is difficult to reconcile with our observations, since simple FM alone does not exhibit EB. To resolve this, we measured the magnetic moment of a \co322 crystal as a function of temperature in both out-of-plane (Fig.~\ref{fig:magnetization}(a)) and in-plane directions (Fig.~\ref{fig:magnetization}(b)). In the out-of-plane direction, a ZFC measurement while warming up in a $10$~mT field shows the expected FM transition at $175$~K, with an additional peak at $125$~K. The same measurement protocol applied in the in-plane direction reveals an AFM-like cusp at $175$~K, followed by a rise in magnetization at $125$~K. The moment amplitude in the ab plane is two orders of magnitude lower than the out-of-plane moment. This is in agreement with other magnetic measurements done previously \cite{Kassem_PRB_2017}. However, the appearance of an in-plane moment with the addition of the $125$~K feature in both orientations points to a previously undiscovered phase transition in \co322. The work by Kassem \textit{et al.} \cite{Kassem_PRB_2017} includes measurements of magnetization and ac susceptibility, which were interpreted as indicating an anomalous magnetic phase preceding the FM phase when cooling. Here, EB clearly shows that the two types of magnetism coexist below $125$~K.
Heat capacity measurements were also performed in order to characterize the $125$~K phase transition. A distinct phase transition at that temperature would be the hallmark of an ordered FM/AFM phase, whereas the lack of such a transition would allude to a glassy one. The results of these measurements are presented in Fig.~\ref{fig:magnetization}(c). A clear transition appears at $175$~K, but no feature is visible at $125$~K, even when following the FC-ZFW protocol mentioned above that shows a transition in transport (as seen in Fig.~\ref{fig:warmups}). This is evidence that the $125$~K transition is not of long range order, but of the freezing of a spin glass.

\section{Discussion}

The classical method for creating EB in materials is to combine the FM material with an AFM one. The AFM is used as a "pinning" layer, as the exchange interaction at the FM-AFM interface pins the FM layer's magnetization, resulting in a higher coercive field needed to flip its magnetic orientation.
It was later discovered that EB can also be induced by a combination of an FM phase and phases other than AFM \cite{Meiklejohn1957}, such as a ferrimagnet \cite{Cain1990} or a spin glass (SG) \cite{Ali_NatMat_2007}. The presence of exchange bias below $T_G = 125$~K, evidenced in Fig.~\ref{fig:cooldownAndEB} and Fig.~\ref{fig:seb}, immediately suggests the coexistence of ferromagnetism with another phase. As we discuss below, we suggest this phase is a dense spin glass arising from strong magnetic frustration in the Kagome lattice.

The first clue as to the nature of the coexisting phase is the unusual bow-tie like structure of the $M(H)$ sweeps in the range $T_G<T<T_c$, shown in Fig.~\ref{fig:biasAtdiffFieldsTemps}. On sweeping the field up, the system begins with a linear response, saturating at a magnetization $M_0$, indicating the polarization of the FM domains. In an ordinary FM, these domains would remain polarized until a sufficiently negative field can flip the spins. In the present case, the domains depolarize at positive fields, and restore their paramagnetic response. As a minimal Landau model, this can be understood as the interplay of two terms in the free energy, one favoring a FM structure $M=\pm M_0$, and another with a minimum favoring $M=0$. This could be the coexistence of an AFM, as claimed recently \cite{Guguchia_arXiv_2019}, or a correlated paramagnet (PM) with AFM interactions. 

The difficulty with reconciling an AFM coexistence at $T_G<T<T_c$, is that EB is not observed in this range, but rather at $T<T_G$. This means that the coexisting phase must become stiffer below $T_G$, not weaker as previously suggested \cite{Guguchia_arXiv_2019}. However, the absence of a heat capacity anomaly at $T_G$, suggests this does not freeze a significant fraction of the degrees of freedom, and in any case muon measurements have confirmed the absence of a competing long range order at low temperatures. 

We suggest that a frustration-driven spin glass transition at $T_G$ can explain all of the present observations. These would lead to exchange bias, and an absent heat capacity anomaly \cite{Ali_NatMat_2007}. This also explains the temperature dependence of the $M(H)$ curves. In the range $T_G<T<T_c$, the system is a FM coexisting with an (antiferromagnetically) correlated PM, leading to a linear susceptibility at low fields, but polarizing at sufficiently high fields. Below $T_G$, the correlated PM freezes into a spin glass, pinning the FM domains into one of two states at $\pm M_0$, causing the hysteresis loops to open up and take the familiar form of a more conventional FM system. This pinning explains the sudden jump in the AHE at $125$~K (Fig. \ref{fig:warmups}) and the opening of the hysteresis loops (Fig. \ref{fig:magnetization}). However, a SG would be expected to show time dependent effects like magnetic relaxation or magnetic memory. We have not observed such effects. This may be because the dynamics of the spin glass are too fast (as expected in dense spin glass) and/or because the magnetization of the FM signal is already so large, that the small magnetic contribution of a SG may be difficult to resolve. 

If the coexisting phase at low temperature is indeed a spin glass, it is unlikely to be disorder driven. All studies of \co322\, have observed the transition to occur at the same temperature. This includes ours and others' samples that are clean enough to show quantum oscillations \cite{Liu_NatPhys_2018, Supplemental}. The spin glass is therefore most likely driven by frustrated interactions on the Kagome lattice. The observation of AFM correlations in Ref.\cite{Guguchia_arXiv_2019} using muon spectroscopy, may be consistent with this interpretation; the FM (with the moment predominantly aligned along c) onsets at $T_C$, however frustrated in-plane AFM correlations cause weak in-plane canting, that subsequently freezes at $T_G$ into a dense SG. The coexistence of the spin glass and the FM phase then leads to exchange biased AHE presently observed.

\section{\label{sec:sum} Summary}
\co322\, is a magnetic Weyl semimetal displaying exchange bias intrinsically, without the need for doping or layering. The exchange bias in this material can also be induced spontaneously at low temperatures and by low magnetic fields. 
We suggest that the origin of the exchange bias is the frustration in the Kagome magnet, which leads to a ferromagnetic state that below $125$~K is simultaneously glassy. The combination of these behaviors leads to exchange bias and spontaneous exchange bias that are strongly evident in the anomalous Hall effect.

We emphasize that magnetism plays in important role in the robustness of the QAHE in magnetically doped topological insulators \cite{Lachman2017}, and therefore may play a crucial role in unlocking the possibility of a QAHE in low-dimensional structures of \co322. The interplay of magnetic frustration and topology in \co322\, provides an example of the potential utility of these materials for future spintronics technologies.

\section{Methods}
\subsection{Single crystal growth}
Single crystals were grown from a stochiometric ratio of elements using the self-flux method (Sn flux). The elements were placed in $AlO_x$ crucible and sealed in an evacuated quartz tube.

\subsection{Transport and magnetization measurements}
Transport measurements were performed in a Quantum Design PPMS, with current flowing in the ab plane.
Mixing was accounted for by removing a constant ratio of $R_{xx}$ from $R_{xy}$ data for all temperatures and fields. This is purely a geometric factor as it is independent of the measurement conditions and is constant for all the transport measurements performed on a specific sample. The ratio was determined to remove ``symmetric" contributions to $R_{xy}$ by ``leveling" the high field values. This protocol was chosen because due to the non-symmetric nature of exchange bias, anti-symmetrising the data was not possible.\\
Magnetization measurements were performed in a Quantum Design MPMS3 on a quartz rod.

\section{Acknowledgements}
\noindent This work was supported by the National Science Foundation under Grant No. 1607753. E.L is an Awardee of the Weizmann Institute of Science - National Postdoctoral Award Program for Advancing Women in Science. 
R.K. is supported by the National Science Foundation (NSF) Graduate Research Fellowship under Grant No. DGE-1106400. 
E.L., N.M. and S.H. acknowledge support from the Gordon and Betty Moore foundation’s EPiQS Initiative through Grant GBMF4374.
High field measurements were performed at the National High Magnetic Field Laboratory, which is supported by National Science Foundation Cooperative Agreement No. DMR-1157490 and the State of Florida.

\section{Author contribution}
E.L performed the crystal growth, transport and magnetization measurements, as well as data analysis for the above measurements.
N.M, R.K and S.H performed high magnetic field measurements.
E.L, R.M and J.A devised the experiments and interpreted the results.   
E.L and J.A wrote the manuscript with contributions from other authors.

\section{Competing interests}
The authors declare no competing interests.

\section{Data availability}
The datasets generated during and/or analysed during the current study are available from the corresponding author on reasonable request.

\bibliography{Co3Sn2S2_EB_18Jun2019}

\end{document}